\documentclass[reprint,showpacs,prl,aps,amsmath,nofootinbib,amssymb,superscriptaddress, notitlepage]{revtex4-1}
\usepackage{mathrsfs} 
\usepackage{natbib}
\usepackage{color}
\usepackage{hyperref}
\usepackage{graphicx}
\usepackage{bm} 
\usepackage{mathrsfs} 
\allowdisplaybreaks
\newcommand{\be}{\begin{equation}}
\newcommand{\ee}{\end{equation}}
\newcommand{\bea}{\begin{eqnarray}}
\newcommand{\eea}{\end{eqnarray}}

\def\spose#1{\hbox to 0pt{#1\hss}}
\def\lta{\mathrel{\spose{\lower 3pt\hbox{$\mathchar"218$}}
          \raise 2.0pt\hbox{$\mathchar"13C$}}}
\def\gta{\mathrel{\spose{\lower 3pt\hbox{$\mathchar"218$}}
          \raise 2.0pt\hbox{$\mathchar"13E$}}}

\begin{document}

\title[The electromagnetic afterglows of gravitational waves as a~test for Quantum Gravity]{The electromagnetic afterglows of gravitational waves as a~test for Quantum Gravity}

\date{07 April 2016} 

\author{M. A. Abramowicz}\email[]{marek.abramowicz@physics.gu.se}
 \affiliation{Nicolaus Copernicus Astronomical Center, ul. Bartycka 18, PL-00-716
               Warszawa, Poland}              
\affiliation{Institute of Physics, Silesian University in Opava,
                   Bezru{\v c}ovo n{\'a}m. 13, CZ-746-01 Opava,
                   Czech Republic}   
                   \affiliation{Department of Physics, University of Gothenburg, SE-412-96 G{\"o}teborg, Sweden}                          
\author{ T. Bulik}
\affiliation{
 Astronomical Observatory, Warsaw University, al. Ujazdowskie 4, PL-00-478 Warszawa, Poland}%
                         \author{ G. F. R. Ellis}
\affiliation{ Mathematics Department, University of Cape Town,
    Rondebosch, Cape Town 7701, South Africa}
     \author{ K. A. Meissner}
\affiliation{ Faculty of Physics, University of Warsaw,
ul. Pasteura 5, PL-02-093 Warszawa, Poland}
    \author{ M. Wielgus}
 \affiliation{Nicolaus Copernicus Astronomical Center, ul. Bartycka 18, PL-00-716
               Warszawa, Poland}              
\

\begin{abstract}
We argue that if particularly powerful electromagnetic afterglows of the gravitational waves bursts will be observed in the future, this could be used as a~strong observational support for some suggested quantum alternatives for black holes (e.g., firewalls and gravastars). A~universal absence of powerful afterglows should be taken as a~suggestive argument against such hypothetical quantum-gravity objects.
\end{abstract}


\maketitle

{\it \textbf{I. Introduction.}} If there is no matter around of a binary system of two uncharged stellar mass black holes, their inspiral-type coalescence (merger) will result in emission of a powerful blast of gravitational waves, but no corresponding electromagnetic radiation, i.e., no  ``afterglow''. Suppose, there is a quantum gravity theory which makes predictions about final stages of stellar evolution very much different than the standard Einstein GR. For example, gravastars or firewalls do form. Then: if gravastars, there are no black holes. So we do not have a reality in which black holes and gravastars may coexist. In this note we point out that coalescence of black hole's quantum-gravity alternatives may yield strong electromagnetic emission with energy $E_\gamma$ comparable to that of the gravitational wave blast itself, $E_{\rm GW}$.  Thus, either standard Einstein and then no afterglows, or something else and then a~possibility for strong afterglows. 

Recently, LIGO reported the first ever detected burst of gravitational radiation (GW150914) from a~black hole merger, that emitted the energy of 
\begin{equation}
E_{\rm GW} \approx 3 \, M_\odot c^2 = 5 \times 10^{54}  {\rm [erg]} .
\end{equation}
Assuming that all of the energy was emitted in the time when the source was observed by LIGO, the duration of the burst was $\Delta t \approx 0.1$[sec], yielding average power
\begin{equation}
    L_{\rm GW} \approx 5 \times 10^{55} {\rm [erg/sec]} = 10^{-4} L_{\rm PLANCK} ,
    \label{Planck}
\end{equation}
    where $L_{\rm PLANCK} = c^5/G$ is the Planck unit for power.
The LIGO and VIRGO teams estimated \cite{abbott16} from a~detailed analysis of the waveform registered during the event that the initial (comparable) masses
of the black holes were $M_1 = 29 \pm 4 M_\odot $ and $M_2 = 36 \pm 4 M_\odot$, and that the final mass was $M_3 = 62 \pm 4 M_\odot$. The analysis was done assuming the validity of standard Einstein's general relativity in a fully dynamical context and neglecting presence of gravity sources other than the merging black holes, i.e., $m \ll M$, where total mass $M \approx M_1 \approx M_2$, where $m$ may represent ordinary matter, but may as well correspond to other forms of energy, such as quantum vacuum.

There were attempts to find an electromagnetic afterglow of GW150914 in gamma rays by the satellite Fermi \cite{connaughton16}.  Certainly, Fermi detected {\it no powerful afterglow} that could be associated with the event GW150914, afterglow with $E_\gamma \approx E_{\rm GW}$ is excluded. Whether there was a~weak afterglow, with $E_\gamma \ll E_{\rm GW}$, is a~matter of debate \cite{Lyutikov16}. Nevertheless, weak afterglows can be explained by the presence of a~small amount of residual matter, $m \ll M \!,$ in the system, e.g., \cite{Loeb16, Stone16}.

The absence of a powerful afterglow associated with the GW150914 event is an important clue. We argue that collisions of firewalls \cite{almhieri13}, gravastars \cite{mazur04}, or other quantum-gravity alternatives to standard Einstein's black holes may result in electromagnetic afterglows of energy $E_\gamma$ comparable to the gravitational waves energy $E_{\rm GW}$. The reason is that such objects are expected to have a mass content $m$ comparable with the total mass, i.e., $m \approx M \! ,$ and a~sizeable fraction of a~corresponding  energy $m c^2$ may be released during the collision event.
\ \\
\ \\
{\it \textbf{II. Firewalls.} } Quantum entanglement of Hawking radiation leads to the the black hole information paradox. One of the suggested remedies for the
paradox supposes the existence of a Planck density $\epsilon_{\rm \small P}$ ``firewall'' with a Planck thickness ${\ell}_{\rm
\footnotesize P}$ near the black hole horizon \cite{almhieri13}. One may estimate the firewall mass to be $m = \epsilon_{\rm \small P} {\ell}^*_{\rm \footnotesize P} {\cal
A}$ by assuming that $m \ll M$. Here ${\cal A} = 16\pi M^2$ is the area of the black hole horizon and
${\ell}^*_{\rm \footnotesize P}$ is a ``proper'' Planck length in the Schwarzschild geometry. This simple calculation has been
done in \cite{abramowicz14}. The result, $m \approx M$, shows that the assumption $m \ll M$ is not correct in this case. Relaxing this assumption and using the standard Einstein field equations (as there are no quantum gravity field equations known, associated with firewalls), \cite{abramowicz14} concludes that {\it Planck density firewalls are
excluded by Einstein's equations for black holes of mass exceeding the Planck mass}. For different reasons, other authors
have also criticized the firewall concept. Susskind's paper on firewalls \cite{susskind12} has been withdrawn with a comment
``the author no longer believes the firewall argument was correct''. Although for many physicists today arguments based on the
standard Einstein equations are not decisive, everyone should accept that {\it independently} on
whether Einstein's field equations are correct or not, {\it if the Planck density firewalls exist}, they should have masses
comparable with masses of their host black holes, $m \approx M$.
\ \\
\ \\
{\it \textbf{III. Gravastars.}} Static, spherically symmetric gravastars models \cite{mazur04} are exact solutions of the standard Einstein field equations. A~gravastar consists of a dark energy sphere, with $p=-\rho c^2 = {\rm const}$ with a radius nearly equal to the gravitational radius $r_{\rm G} = 2MG/c^2$, surrounded by a shell of matter with the extreme equation of state $p = \rho c^2$. For ``standard'' gravastars \cite{mazur04} the shell thickness is of the order of the Planck lenght, $\Delta \approx \ell_P$, but recently non-standard gravastars have been constructed \cite{Chirenti16} with much thicker shells, $\Delta \gg \ell_P$. Outside the gravastar, $r > r_{\rm G}$, the metric is that of the standard vacuum Schwarzschild solution. 

As an illustration of a specific possible mechanism leading to emission of an energetic afterglow, we roughly estimate how much of electromagnetic radiation is emitted when a~ball of a~superconducting dark matter $p=-\rho c^2$ is almost instantaneously squeezed (this mimics a~``gravastar merger''). A~gravastar of a~total mass $M = 30 M_\odot$ has a $9\times 10^6$ [cm] radius, the density is equal to $\rho \approx 2 \times 10^{13}\ {\rm [g/cm^3]}$. Energy density in the vacuum in a~theory of a~complex scalar field $\phi$ with spontaneously broken $U(1)$ symmetry is approximately equal to 
\begin{equation}
\rho\approx \frac{\hbar}{c}\lambda\phi_0^4 ,
\end{equation}
where $\lambda$ is the self-coupling constant of the field and $\phi_0$ is its (real) vacuum expectation value. Assuming $\lambda\approx 1$ we get $ \phi_0\approx 5 \times 10^{12} \ {\rm [1/cm]} $. The vector potential ${\bf A}$ for a~superconducting sample is non vanishing only for a~small penetration length $\delta$ near the surface (Meissner effect).
From the London equations we find
\begin{equation}
\mu_0 {\mathbf j}=-\frac{1}{\delta^2} {\mathbf A} .
\end{equation}
The current density is proportional to the derivative of the field times the field so
\begin{equation}
j\approx ec\frac{\phi_0^2}{\delta} ,
\end{equation}
where $e$ is the electron charge, and hence
\begin{equation}
A\approx \delta\mu_0 ec\phi_0^2 .
\end{equation}
Magnetic field is a rotation of ${\mathbf A}$, hence
\begin{equation}
B\approx \mu_0 e c\phi_0^2 .
\end{equation}
Its value does not depend on $\delta$ but it is non-zero only in the shell of thickness $\delta$.
If we squeeze the sphere then there is also an electric field of magnitude $B v_d$ where $v_d$ is the velocity of the deformation. Assuming that the Lorentz factor $\gamma$ of the star rotation is not too big (for the event GW150914 velocities were around $0.6 c$ so $\gamma\approx 1.25$), we can neglect $\gamma^6$ factor and we find the power (Poynting vector times the area)
\begin{equation}
L_\gamma  = \frac{{\rm d}E_\gamma}{{\rm d} t }\approx \frac{B^2 S_{\rm eff} v_d}{\mu_0  } ,
\end{equation}
where $S_{\rm eff}$ is the effective area (some fraction of the total area).
Gathering all the results we get the total power
\begin{equation}
L_\gamma \approx  \mu_0 e^2 c^2 \phi_0^4 S_{\rm eff} v_d \approx \alpha_{\rm em}\frac{S_{\rm eff}}{S}\frac{v_d}{c} L_{\rm PLANCK}.
\end{equation}
Assuming $S_{\rm eff}\approx 0.1 S$ and estimating $v_d$ as $R/\tau\approx 10^{-2} c$, where $\tau$ is the total duration of the event, we get
\begin{equation}
L_\gamma \approx 10^{-4} L_{\rm PLANCK} \approx 10^{55} {\rm [erg/sec]} .
\label{Lgamma}
\end{equation}
This is of the same order as the estimated peak of the gravitational wave power in the event GW150914. Note that both $L_{\rm GW}$, given by Eq. (\ref{Planck}), and $L_\gamma$, given by Eq. (\ref{Lgamma}), correspond to fixed fractions of $L_{\rm PLANCK}$ and therefore they do not scale with the total mass $M$; collisions of supermassive binaries of black holes or gravastars would result in $L_{\rm GW}$ and $L_\gamma$ on the same order of magnitude as collisions of the stellar mass objects. A~non-standard gravastars, with a thicker $p =\rho c^2 $ shell has a~smaller amount of dark energy to produce the afterglow via the mechanism considered above, but more ordinary matter $p = \rho c^2 $ to produce an afterglow via standard radiative processes. Indeed, for a sufficiently thick shell there could be more afterglow energy coming from the  $p = \rho c^2$ part than from the $p = -\rho c^2$ part of the gravastar [Rezzolla, private communication]. Therefore, gravastars (standard or not) may have  $mc^2 \approx Mc^2$
available for an electromagnetic afterglow. However, determining how large the electromagnetic emission would be at infinity is far from trivial: a~common apparent horizon may be formed before the two gravastars enter in contact \cite{Jaramillo12}. This may, in principle, prevent colliding gravastars from having any observable electromagnetic emission.
\ \\
\ \\
{\it \textbf{IV. Other possibilities.}} When field equations of a particular quantum gravity theory are known, it is of course (in principle) possible to calculate, in a~very detailed way, all observational consequences of collisions of ``black holes'' predicted in the theory. This is the case of
Ho{\v r}ava's quantum gravity \cite{horava09}; see a~list of relevant references in \cite{Barausse11} and \cite{goluchova15}. Calculating Ho{\v r}ava's black hole ringdown may be particularly interesting in the view of the recent argument that the gravitational ringdown
may not be a probe of the event horizon \cite{Barausse14, cardoso16}; see, however, \cite{Chirenti16}. One of the arguments discussed in \cite{cardoso16} in the context 
of the gravitational radiation, resembles what was pointed out in \cite{abramowicz02}, namely that there is no observational
proof possible for the existence of the event horizon, based on electromagnetic radiation. These observations may probe only the
existence of circular light ray (at $r = 1.5 \,r_{\rm G}$ in the Schwarzschild spacetime) but not smaller radii. A~somehow exotic possibility of a~powerful electromagnetic afterglow due to a~collision of Ho{\v r}ava's naked singularities was discussed in \cite{Malafarina16}.
\ \\
\ \\
{\it \textbf{ V. Conclusions.}} 
We wish to conclude that detection of powerful afterglows will provide an observational support for the existence of quantum alternatives for black holes, but a~universal absence of such afterglows will constitute an argument against them. Our point that collisions of standard black holes (with ``pure'' horizons) are much dimmer in electromagnetic radiation than collisions of non-standard quantum black holes (with no horizon or with a ``dirty'' horizon), resembles arguments advocated by Lasota, Narayan and others \cite{menou99}, \cite{narayan03} that accreting black holes are much dimmer than similarly accreting neutron stars.
\ \\
\ \\
{\it \textbf{ Acknowledgments.}} We thank A. R\'o\.za\'nska, W.~Klu\'zniak, J.-P. Lasota, E.~Barausse, V. Cardoso and L.~Rezzolla for their help and comments. KAM thanks Albert Einstein Institute in Potsdam for hospitality and support. MW acknowledges support of the Foundation for Polish Science within the START Programme.

\end{document}